\documentclass[12pt]{article}
\usepackage{graphics, color}
\usepackage{graphicx}
\usepackage{amssymb}

\def\gsim{\, \rlap{$>$}{\lower 1.1ex\hbox{$\sim$}}\,}
\def\lsim{\, \rlap{$<$}{\lower 1.1ex\hbox{$\sim$}}\,}

\begin{document}


\begin{titlepage}
\bigskip
\bigskip\bigskip\bigskip
\centerline{\Large Cosmic String Loops, Large and Small}
\bigskip\bigskip\bigskip
\bigskip\bigskip\bigskip

\centerline{{\bf Florian Dubath}\footnote{\tt dubath@kitp.ucsb.edu}}
\medskip
 \centerline{{\bf Joseph Polchinski}\footnote{\tt joep@kitp.ucsb.edu}}
\medskip
\centerline{\em Kavli Institute for Theoretical Physics}
\centerline{\em University of California}
\centerline{\em Santa Barbara, CA 93106-4030}\bigskip
\bigskip
\centerline{{\bf Jorge V. Rocha}\footnote{\tt jrocha@physics.ucsb.edu}}
\medskip
\centerline{\em Department of Physics}
\centerline{\em University of California} \centerline{\em Santa Barbara, CA 93106}
\bigskip
\bigskip
\bigskip\bigskip


\begin{abstract}
We extend our earlier model of the small scale structure of cosmic string networks through an improved treatment of the separation of long and short scales.  We find that the production of small loops (at the gravitational radiation scale) is a robust feature of string networks, in addition to a population of loops near the horizon scale.  We obtain quantitative agreement with the scaling of loop production functions as found in simulations by two groups.
\end{abstract}
\end{titlepage}
\baselineskip = 16pt


Estimates of the size of cosmic string loops are uncertain to many orders of magnitudes, ranging from near-horizon scales~\cite{Kibble:1984hp,Martins:2005es,Vanchurin:2005pa,Olum:2006ix} 
down to the gravitational radiation length scale~\cite{Bennett:1989ak,Siemens:2001dx,Ringeval:2005kr,Polchinski:2007rg}, and even to microscopic scales~\cite{Vincent:1996rb}.  Even when the properties of the string are specified, so that the problem is completely well-posed, this uncertainty arises from the inability to solve the equations for the evolution of the network.  Numerical simulations have been unable to reach the necessary dynamical ranges of length and time, while analytic methods 
have difficulty with the nonlinearities of the problem.

There has recently been progress in both approaches.  By using a doubling trick, recent simulations have been able to reach a much larger dynamical range~\cite{Vanchurin:2005pa,Olum:2006ix}.  By focusing on the properties of the network well below the horizon scale, it has been possible to formulate an analytic model that captures the quantitative properties of the short distance structure~\cite{Polchinski:2006ee}.  In this paper, we will argue that these two recent advances point to a common picture.

The new simulations, taken at face value, show two peaks in the loop production function~\cite{Vanchurin:2005pa,Olum:2006ix}.  One is at a surprisingly large scale, around a tenth of the horizon length, while the other is near the UV cutoff.  The large loops have strong gravitational wave signatures, and so lead to stronger bounds and enhanced prospects for future sensitivity~\cite{Olum:2006at,Hogan:2006we,Siemens:2006yp}.  However, the total length going into loops is fixed by energy conservation, so any string that goes into small loops reduces this effect~\cite{Polchinski:2007qc}.  The simulations at their current dynamic ranges show roughly 90\% of the string going into small loops, but the authors of that work conjecture that these loops are transient, so that in final attractor solution essentially all string will go into long loops.

The analytic model begins by considering the small scale structure on long strings~\cite{Polchinski:2006ee}.  It is found that the string approaches fractal dimension 1 at short distances, with a deviation that goes as a power of the separation $l$,
\begin{equation}
d_f = 1 + O (l/t)^{2\chi}\ .
\label{fracdim}
\end{equation}
The exponent $\chi$ is determined in terms of the rate of expansion of the universe, $a(t) \propto t^{\nu}$ and the mean velocity squared $v^2$ in the network, taken from simulations:
\begin{equation}
 \chi = \frac{\nu ( 1 - 2v^2)}{1 - \nu ( 1 - 2v^2)}\ .
\end{equation}
The relevant values are
\begin{eqnarray}
{\rm radiation\ era:}&& \nu = \frac{1}{2}\ ,\quad v^2 = 0.41\ ,\quad \chi = 0.10\ ,
\nonumber\\
{\rm matter\ era:}&& \nu = \frac{2}{3}\ ,\quad v^2 = 0.35\ ,\quad \chi = 0.25\ .
\end{eqnarray}
The predicted two-point function appears to agree with simulations~\cite{Martins:2005es,HS}.
Taking this structure as input, we can then calculate the rate at which self-intersections of the long string take place to produce a loop, as a function of loop size.  One finds that, even though the string is becoming straight at short distance, the loop production function (that is, the rate of loop production weighted by loop length) diverges at small lengths~\cite{Polchinski:2006ee}.  This is because the exponent $\chi$ is rather small in both eras, meaning that the approach to fractal dimension 1 is slow: the critical value for small loop production is $\chi = 0.5$.\footnote{In Ref.~\cite{Martins:2005es} the loop production function shows no evidence of this divergence but this is expected: in those simulations the short distance structure seems to change below a certain length scale in such a way that $\chi$ exceeds the critical value.  This could be an artifact of the finiteness of their dynamic range.}

The divergence in the loop production function must be cut off, because the total rate at which string length goes into loops is fixed by energy conservation.  In the approach of Ref.~\cite{Polchinski:2006ee}, this condition saturated at a length scale not too far below the horizon.  It was then argued that smaller loops would be produced by extensive fragmentation, but this appeared to be highly nonlinear and not amenable to analytic calculation.  In the present paper we make an essential improvement on the previous model, which leads to a very different, and simpler, picture.

On the time scale of small loop formation, we can ignore the expansion of the universe.  The string configuration is described by the left- and right-moving unit tangent vectors
\begin{equation}
({\bf p}_+(u), {\bf p}_-(v)) = 2 \left( \partial_u {\bf x}(u,v), \partial_v {\bf x}(u,v) \right)\ ,
\quad
(u, v) = (t + \sigma,\, t - \sigma)\ .
\end{equation}
The calculation of the loop production rate requires separating the ${\bf p}_\pm$ into a long-distance piece and a short-distance piece.  In Ref.~\cite{Polchinski:2006ee} the long-distance part was simply averaged over the unit sphere, with a weight factor chosen to give the correct mean value of ${\bf p}_+ \cdot {\bf p}_-$.  In the present work we will take a long-distance part with some given classical $(u, v)$-dependence, and then average over the short distance part with a weighting corresponding to Eq.~(\ref{fracdim}).  In the end we should average over an ensemble of long-distance configurations, but the result will be largely independent of this.  The inclusion of the time-dependence will reveal that loop formation leads to a strong correction to the distribution of long-distance configurations, which was not accounted for in the earlier approach~\cite{Polchinski:2006ee}.\footnote{These same issues were faced in the pioneering work~\cite{Austin:1993rg}, in a somewhat different framework.}

A typical configuration is shown in Fig.~1.
\begin{figure}
\begin{center}
\ \includegraphics[width=19pc]{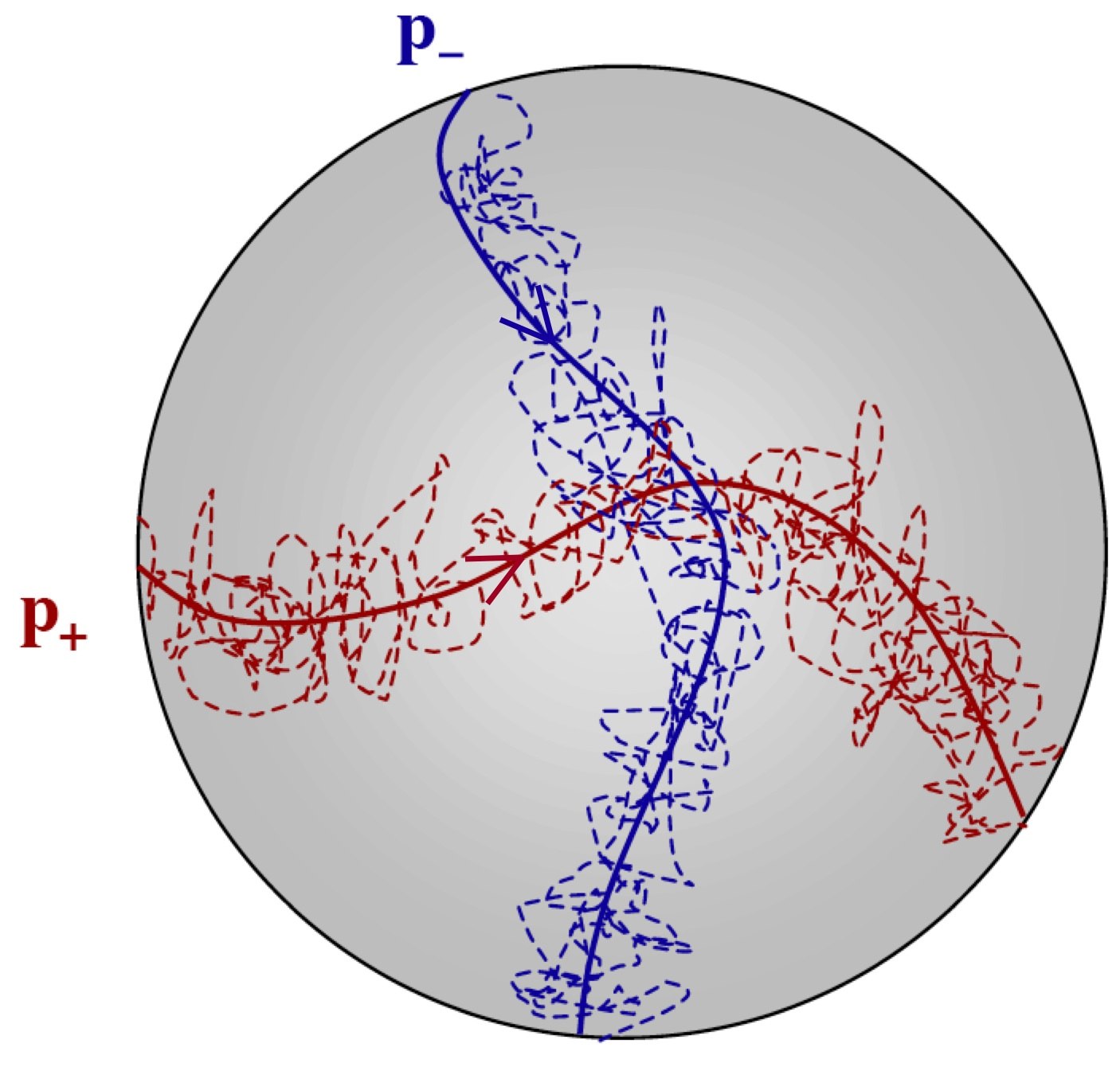}
\end{center}
\caption{The functions ${\bf p}_\pm$ written as a fixed long-distance piece (bold) plus a random short-distance part.  The values of $u$ and $v$ increase along the arrows, with the classical cusp defined to lie at $(u,v) = (0,0)$.}
\end{figure}
The long-distance configuration is in bold, and the total configuration is dashed.  Note that, while the string configuration approaches fractal dimension 1 at short distance, the graph of its tangent has large fractal dimension $1/\chi$~\cite{Polchinski:2006ee}.  

In the figure, we have shown a long distance configuration in which the curves ${\bf p}_+(u)$ and ${\bf p}_-(v)$ cross, meaning that there is a cusp in the spacetime evolution of the string~\cite{Turok:1984cn}.  It is evident that the effect of the short distance structure is to turn this one large cusp into many small cusps.  The formation of small loops is closely associated with the cusps.  The condition for a cusp is crossing of the two curves at a point~${\bf p}_+(u) = {\bf p}_-(v)$.  The condition for self-intersection of the string is
\begin{eqnarray}
{\bf L}_+(u,l) &=& {\bf L}_-(v,l)\ ,
\nonumber\\
{\bf L}_+(u,l) &\equiv& \int_{u-l/2}^{u+l/2} du'\, {\bf p}_+(u') \ , \quad
{\bf L}_-(v,l) \equiv \int_{v-l/2}^{v+l/2} dv'\, {\bf p}_-(v') \ ;
\end{eqnarray}
when this is satisfied the segment of length $l$, centered at $(u,v)$, breaks off as a loop.  For small $l$ the functions ${\bf p}_\pm$ will have small variation along the segment, so small loops will form where these are approximately equal~\cite{Albrecht:1989mu,SiemOlum2003}.

Following this discussion, it is clear that at early times, when $u$ and/or $v$ are sufficiently negative, the curves will be outside the cusp region and the production of small loops will be negligible.  As the cusp begins to form, production of small loops begins to increase.  We will find that loops of all scales begin to form at the same time, extending all the way down to the gravitational radiation cutoff.  This is in marked contrast to the cascade from large to small loops, anticipated in Ref.~\cite{Polchinski:2006ee}.  Thus we are led to conclude that the production of loops down to the gravitational radiation scale is inevitable, and not a transient effect.

The production of large loops is outside our model, but these also seem inevitable.  In the course of the evolution of the network it must happen at least occasionally that a large loop will form and survive reconnection.  We must leave it to the simulations to determine the relative weight and average size of these loops, but we can understand why they do not fragment entirely into much smaller loops.  From Fig.~1 we expect that if there are any cusps on these loops then the regions containing them will dissolve into small loops.  However, if there are parts of the loop where ${\bf p}_+$ is sufficiently far from the ${\bf p}_-$ curve, and vice versa, then these will survive.  Thus we expect that if the functions ${\bf p}_\pm$ are plotted, there will be significant gaps in each, corresponding to a large kink, and the surviving curves will avoid each other.  It seems plausible that the total amount of string surviving in the long loops is a small fraction of the total, as seems to be true with the current dynamical range~\cite{Vanchurin:2005pa,Olum:2006ix}, but it is important to study this further.

The loop production cuts off when the probability for a given bit of string to be incorporated in a loop reaches 1.  We will see that this occurs long before the classical cusp point $(u,v)=(0,0)$ is reached.  Thus, whereas in Ref.~\cite{Polchinski:2006ee} the region where the long-distance parts of ${\bf p}_+(u)$ and ${\bf p}_-(v)$ were parallel gave the dominant contribution, but in a correct treatment it is simply absent.  This explains why this earlier treatment gave a normalization for the loop production function that was much too large, though it did seem to find the correct functional form.

In the remainder of this paper we will carry out the calculation just described.  As further support for our picture, we will obtain a prediction for the small loop production function that fits well with the simulations by two groups.  The fit actually works better than anticipated, as there is no sign that it is distorted by further fragmentation; we speculate as to why this might be.

To carry out the calculation of the rate, we separate the left- and right-moving unit tangent vectors into their long- and short-distance parts,
\begin{equation}
{\bf p}_\pm = {\bf l}_{\pm} + {\bf s}_\pm\ .
\end{equation}
Since the production of small loops takes place near the cusp, we take the simplest cusp form for the long-distance structure,
\begin{eqnarray}
{\bf l}_+ &=&  \hat {\bf z}  (1 - V_+^2 u^2)^{1/2}+ {\bf V}_+ u \approx  \hat {\bf z} + {\bf V}_+ u -\frac{\hat {\bf z} }{2}  V_+^2 u^2
\ ,
\nonumber\\
{\bf l}_- &=&  \hat {\bf z}  (1 - V_-^2 v^2)^{1/2}+ {\bf V}_- v \approx  \hat {\bf z} + {\bf V}_- v -\frac{\hat {\bf z} }{2}   V_-^2 v^2
\end{eqnarray}
with ${\bf V}_+ \cdot \hat {\bf z} = {\bf V}_- \cdot \hat {\bf z} = 0$.  We are expanding to order $u^2$, $v^2$.
The slopes ${\bf V}_{\pm}$ have units of inverse length, and are reciprocal to the size of the cusp.  They will be of order the inverse correlation length, somewhat less than the horizon length.

Adding in the short distance structure gives, to the same order,
\begin{eqnarray}
{\bf p}_\pm &=& \hat {\bf z} + {\bf y}_\pm - \frac{\hat {\bf z}}{2} y_\pm^2  \ ,\nonumber\\
{\bf y}_+(u) &=& {\bf V}_+ u + {\bf w}_+(u)\ ,\quad {\bf y}_-(v) = {\bf V}_- v + {\bf w}_-(v) \ ,
\end{eqnarray}
with
\begin{eqnarray}
\langle {\bf w_+}(u) \cdot {\bf w_+}(u') \rangle
&=&
 {\cal A} c_{\chi} \int_{-\infty}^\infty \frac{dk}{2\pi} e^{i k (u-u')} \frac{1 }{ |k|^{1+ 2\chi} t^{2\chi}}
(e^{-|k| l_{\rm GW}} - e^{-|k| L}) \nonumber\\
&=& \frac{\cal A}{t^{2\chi} \cos \pi\chi}  {\rm Re}\!\left[ (i |u - u'| + L)^{2\chi}  - (i |u - u'| +  l_{\rm GW})^{2\chi} \right]\ , \label{2pf}
\end{eqnarray}
and similarly for ${\bf w}_-$; here $c_\chi = 2\sin(\pi\chi) \Gamma(1+2\chi)$ and ${\cal A}$ is simply related to the normalization of the $O(l/t)^{2\chi}$ correction in equation~(\ref{fracdim}).  It is approximately equal to $0.6$ during both radiation and matter domination~\cite{Polchinski:2006ee}.  We have introduced a smooth short-distance cutoff at the scale $l_{\rm GW} \sim {\cal A} (G\mu)^{1 + 2\chi} t$~\cite{Polchinski:2007rg} to take account of the effects of gravitational radiation.  The cutoff $L$ represents the separation between the long-distance and short-distance parts of the configuration, taken around the correlation length.  
In the range $l_{\rm GW} \ll |u-u'| \ll  L$ expression~(\ref{2pf}) reduces to
\begin{equation}
\langle {\bf w_+}(u) \cdot {\bf w_+}(u') \rangle \approx {\cal A}\, (L'^{2\chi}  - |u - u'|^{2\chi} )/t^{2\chi}\ , \label{2pfa}
\end{equation}
with $L'^{2\chi} = L^{2\chi}/{\cos \pi\chi}$; the $u, u'$ dependence is as in Ref.~\cite{Polchinski:2006ee}.

For  reconnection probability unity, which we assume, the rate of loop formation is given by the rate of self-intersection,
\begin{equation}
d{\cal N} = \left\langle \delta^{3}({\bf L}_+(u,l) - {\bf L}_-(v,l)) \left| \det {\bf J} (u,v,l)\right |  \right\rangle du\, dv\, dl \ ,
\label{dN}
\end{equation}
where ${\bf L}_\pm$ were defined above and the Jacobian is
\begin{equation}
{\bf J}(u,v,l) = \left[ \begin{array}{c} 
{\bf p}_+(u+l/2) - {\bf p}_+(u-l/2)\\
 {\bf p}_-(v+l/2) - {\bf p}_-(v-l/2)\\
\frac{1}{2}{[{\bf p}_+(u+l/2) + {\bf p}_+(u-l/2) - {\bf p}_-(v+l/2) - {\bf p}_-(v-l/2)] }
\end{array}
\right]\ .
\end{equation}
In our previous work~\cite{Polchinski:2006ee} we were able to estimate the expectation value of the product~(\ref{dN}) by the product of expectation values, but here there are strong correlations: when the delta-function is nonzero, the third row of the Jacobian is much smaller than its mean value.  We therefore go to new variables, separating ${\bf w}_\pm$ into a piece constant on the segment and a piece with zero average on the segment,
\begin{eqnarray}
{\bf w}_+(u') &=& {\bf W}_+ + {\mbox{\boldmath$ \omega$}}_+(u')\ , \quad {\bf W}_+ = \frac{1}{l} \int_{u-l/2}^{u+l/2} du'\, {\bf w}_+(u') \ ,\nonumber\\
{\bf w}_-(v') &=& {\bf W}_- + {\mbox{\boldmath$ \omega$}}_-(v')\ , \quad {\bf W}_- = \frac{1}{l} \int_{v-l/2}^{v+l/2} dv'\, {\bf w}_-(v') \ .
\end{eqnarray}
The variables ${\mbox{\boldmath$ \omega$}}_\pm$ and ${\bf W}_\pm$ depend on the parameters $u, v, l$ of the loop, but we leave this implicit.  We will be interested in the loop production in the range $l_{\rm GW} \ll l \ll  L$, so we use the form~(\ref{2pfa}).
One then finds the two-point functions
\begin{eqnarray}
t^{2\chi}\langle {{\mbox{\boldmath$ \omega$}}_+}(u') \cdot {{\mbox{\boldmath$ \omega$}}_+}(u'') \rangle
&=& - {\cal A} |u' - u''|^{2\chi} + f(u') + f(u'') \ ,\label{oo}
\\
t^{2\chi}\langle {{\mbox{\boldmath$ \omega$}}_+}(u') \cdot {\bf W_+} \rangle
&=& - f(u') \ ,
\label{cross}\\
t^{2\chi}\langle {\bf W}_+ \cdot {\bf W_+} \rangle &=& {\cal A} L'^{2\chi} + O(l^{2\chi})\ , \label{WW}
\end{eqnarray}
where 
\begin{equation}
f(u') = \frac{{\cal A}}{(2\chi+1) l} \left[ (l/2 + u' - u)^{2\chi+1} + (l/2 - u' + u)^{2\chi+1} - \frac{l^{2\chi+1}}{2\chi+2} \right]  
\label{fsigma}
\end{equation}
is defined only for points on the segment, i.e. $u-\frac{l}{2} \leq u' \leq u+\frac{l}{2}$.

Now let us express the rate in terms of these quantities.  First, for the transverse parts of ${\bf L}_\pm$ we have
\begin{equation}
{\bf L}^\perp_+ = l({\bf W}_+ + {\bf V}_+ u)\ ,\quad 
{\bf L}^\perp_- = l({\bf W}_- + {\bf V}_- v) \ .
\end{equation}
The transverse part of the delta-function sets these equal, ${\bf L}^\perp_+ = {\bf L}^\perp_-  \equiv {\bf L}^\perp$, and we use this in evaluating the remaining terms.
For the $z$ component,
\begin{eqnarray}
{L}^z_+ &=& l - \frac{1}{2} \int_{u-l/2}^{u+l/2} du' \left[ {\bf V}_+ u'  + {\bf W}_+ + {\mbox{\boldmath$ \omega$}}_+ (u') \right]^2 \nonumber\\
&=& l - \frac{1}{2l} {{\bf L}^\perp}^2  - \frac{1}{2} \int_{u-l/2}^{u+l/2}  du' \left[ {\mbox{\boldmath$ \omega$}}_+ (u') + {\bf V}_+ (u' - u) \right]^2 \ ,
\end{eqnarray}
where we have used the fact that the mean value of ${\mbox{\boldmath$ \omega$}}_+$ is zero.  In the integrand, the fluctuations of ${\mbox{\boldmath$ \omega$}}_+$ are of order $l^\chi$, while the classical term is of order $l$, so we can drop the latter for small loops.  Thus,
\begin{equation}
{L}^z_+ - {L}^z_- = - \frac{1}{2} \int_{u-l/2}^{u+l/2} du' \left[ {\mbox{\boldmath$ \omega$}}_+ (u') \right]^2
+ \frac{1}{2} \int_{v-l/2}^{v+l/2} dv' \left[ {\mbox{\boldmath$ \omega$}}_- (v') \right]^2 = O(l^{1+2\chi})\ .
\end{equation}
For the transverse part of the first row of ${\bf J}$,
\begin{eqnarray}
{\bf p}^\perp_+(u+l/2) - {\bf p}^\perp_+(u-l/2) &=& {\mbox{\boldmath$ \omega$}}_+(u+l/2) - {\mbox{\boldmath$ \omega$}}_+(u-l/2) + {\bf V}l  \nonumber\\
 &=& {\mbox{\boldmath$ \omega$}}_+(u+l/2) - {\mbox{\boldmath$ \omega$}}_+(u-l/2) + 
 O(l)\ .
 \end{eqnarray}
Similarly in the second and third rows, in the transverse terms we can replace ${\bf p}_\pm$ with $ {\mbox{\boldmath$ \omega$}}_\pm$, after imposing ${\bf L}^\perp_+ = {\bf L}^\perp_-$.  In the $z$ component of the first row,
\begin{eqnarray}
{ p}^z _+(u+l/2) - {p}^z_+(u-l/2) &=& - \frac{1}{l} {\bf L}^\perp \cdot [ {\mbox{\boldmath$ \omega$}}_+(u+l/2) - {\mbox{\boldmath$ \omega$}}_+(u-l/2)] \nonumber\\
&&\quad
- \frac{1}{2} \Bigl( [{\mbox{\boldmath$ \omega$}}_+(u+l/2)]^2 - [{\mbox{\boldmath$ \omega$}}_+(u-l/2)]^2 \Bigr) + O(l^{1+\chi})\ .\quad \label{jz1}
 \end{eqnarray}
The first term is of order $l^\chi$ and the second of order $l^{2\chi}$, but the first actually drops out.  In all three rows of ${\bf J}$ one finds the same pattern, $J^z = -{\bf L}^\perp \cdot {\bf J}^\perp + J^z[{\mbox{\boldmath$ \omega$}}]$.  Thus the first term in the $z$-column is linearly dependent on the other two columns, and drops out in the determinant, leaving $J^z$ with ${\bf p}_\pm$ replaced by ${\mbox{\boldmath$ \omega$}}$.

We have now expressed the terms multiplying $\delta^{2}({\bf L}^\perp_+ - {\bf L}^\perp_- )$ all in terms of ${\mbox{\boldmath$ \omega$}}$.  The cross-correlation~(\ref{cross}) is of order $l^{2\chi}$, smaller than the geometric mean $L'^\chi l^\chi$ of the diagonal correlators~(\ref{oo},$\,$\ref{WW}).  We can therefore ignore it, giving
\begin{equation}
\left\langle \delta^{3}({\bf L}_+ - {\bf L}_- ) \left| \det {\bf J} \right |  \right\rangle 
= \left\langle \delta^{2}({\bf L}^\perp_+ - {\bf L}^\perp_- ) \rangle \times \langle
\delta({L}^z_+ - {L}^z_-) 
\left| \det {\bf J} \right |  \right\rangle_{ {\bf p}_\pm  \to {\mbox{\boldmath\scriptsize$ \omega_\pm$}}}\ .
\end{equation}
Thus we have achieved our aim of factorizing the expectation value.  In the second term the delta-function scales as $l^{-1-2\chi}$, inversely to its argument, while the columns of ${\bf J}$ scale as $l^\chi, l^\chi, l^{2\chi}$, giving the same overall $l^{-1+2\chi}$ scaling as in the previous work~\cite{Polchinski:2006ee}.  In fact, this second term is identical to that in Ref.~\cite{Polchinski:2006ee}.  It was evaluated there with the assumption that the short-distance structure is gaussian, giving a numerical coefficient of $2.2 {\cal A}$ in the radiation era and $1.3 {\cal A}$ in the matter era.  As we will discuss below, the gaussian approximation may not be valid here, but this does not alter the scaling with $l$.

It remains to evaluate
\begin{eqnarray}
\left\langle \delta^{2}({\bf L}^\perp_+ - {\bf L}^\perp_- ) \right\rangle
&=&
\int \frac{d^2 q}{(2\pi)^2} \left\langle e^{i {\bf q} \cdot ({\bf L}^\perp_+ - {\bf L}^\perp_- ) }\right\rangle
\nonumber\\
&=& \int \frac{d^2 q}{(2\pi)^2} e^{- {\cal A} L'^{2\chi} l^2 q^2/4 t^{2\chi} + i l {\bf q}\cdot ({\bf V}_+ u - {\bf V}_- v)}
\nonumber\\
&=& \frac{t^{2\chi} e^{-({\bf V}_+ u - {\bf V}_- v)^2 t^{2\chi}/{\cal A} L'^{2\chi}}}{\pi {\cal A} L'^{2\chi} l^2} \ .
\end{eqnarray}
In all,
\begin{equation}
d{\cal N} = C L'^{-2\chi} e^{-({\bf V}_+ u - {\bf V}_- v)^2 t^{2\chi}/{\cal A} L'^{2\chi}} \, du\, dv\, \frac{dl}{ l^{3 - 2\chi}}\ ,
\label{dn} 
\end{equation}
where (with the gaussian approximation) $C = 0.7$ in the radiation era and $C = 0.4$ in the matter era.  These constants are actually independent of ${\cal A}$ so the normalization of the two-point function only enters the calculation of loop production through the exponential suppression factor.

\begin{figure}
\centering
\includegraphics[width=33pc]{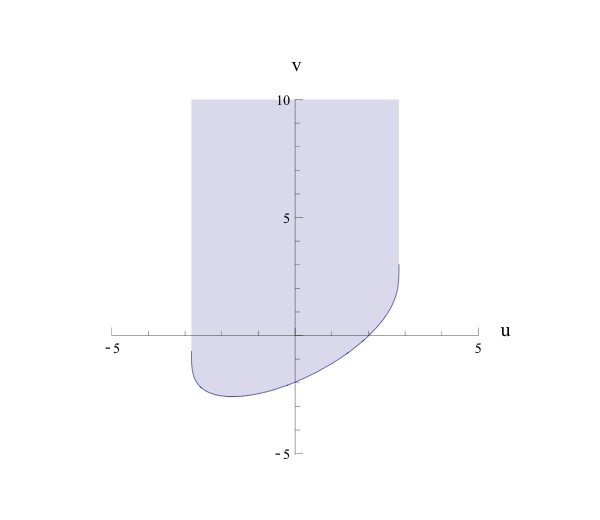}
\caption{The curve ${\bf p_+}$ is excised in the shaded region of the $(u,v)$-plane by loop formation.  The cusp occurs at $u=v=0$ and the curve that delineates the excised region is $v_{max}(u)$.  Both $u$ and $v$ are expressed in units of the cosmological time $t$.  For this particular example the values $G\mu=10^{-9}$ and $\cos(\theta) = 2/\pi$ were used, where $\theta$ denotes the angle between ${\bf V}_+$ and ${\bf V}_-$.}
\label{excision}
\end{figure}

Now consider a left-moving point with given $u$.  The total probability per unit $v$ that this point be incorporated into a loop is
\begin{equation}
\frac{d{\cal P}}{dv} = C L'^{-2\chi}e^{-({\bf V}_+ u - {\bf V}_- v)^2 t^{2\chi}/{\cal A} L'^{2\chi}} \int \frac{dl}{ l^{2 - 2\chi}} \ .
\end{equation}
Note that we have replaced $\int du \to l$ to count the loops containing the given point.  This diverges at small $l$~\cite{Polchinski:2006ee}, because we have not yet taken into account the smoothing due to gravitational radiation in the two-point function~(\ref{2pf}).  The smoothed two-point function is quadratic at $u' - u \to 0$, corresponding to the form~(\ref{2pfa}) at $\chi = 1$: at the shortest distances the divergence is gone (see also Ref.~\cite{Rocha:2007ni}).  Thus we cut the integral off at $l_{\rm GW}$ to get
\begin{equation}
\frac{d{\cal P}}{dv} \approx  l_{\rm GW}^{-1 + 2\chi}  L'^{-2\chi} e^{-({\bf V}_+ u - {\bf V}_- v)^2 t^{2\chi}/{\cal A} L'^{2\chi}} \ . \label{dpdv}
\end{equation}
At early times, where $v$ is large, the probability of a loop containing $u$ is small due to the gaussian.
However, $l_{\rm GW}^{-1 + 2\chi}$ is large compared to the other dimensionful quantities, so as we integrate in $v$ we soon reach total probability $1$ for a range of values of $u$ near the cusp: this portion of the string is removed by loop production (see Fig.~\ref{excision}).  Therefore, for each value of the $u$-coordinate, the maximum value of $v$ that may still not be included in a loop is the solution (when it exists) of the equation ${\cal P}(v_{max})=1$, where
\begin{equation}
{\cal P}(v(u)) = \frac{C}{(1-2\chi)l_{GW}} \left(\frac{l_{GW}}{L'}\right)^{2\chi} 
  \int_{-\infty}^{v(u)} dv \, e^{-({\bf V}_+ u - {\bf V}_- v)^2 t^{2\chi}/{\cal A} L'^{2\chi}}  \ .
\end{equation}
We see that the excised region depends on the the angle $\theta$ between ${\bf V}_+$ and ${\bf V}_-$ for each cusp, not to mention the ubiquitous $G\mu$.  To obtain the normalization of the loop production function would require the exact knowledge of the probability distribution for the angles $\theta$ as well as the density of cusps.  One can certainly obtain normalizations in accordance with the simulations but there are too many free parameters so the answer is inconclusive.

These are our main results.  First, loops of all sizes down to the UV cutoff form simultaneously, rather than in a cascade of fragmentation.  Therefore the production of the small loops is a robust physical result.  The integral of the probability~(\ref{dpdv}) for a point on the string to break off becomes large long before the cusp $v=0$ is reached, so the overcounting in Ref.~\cite{Polchinski:2006ee} is explained.  However, this different treatment of the long-distance structure does not affect the short distance part of the rate~(\ref{dn}), so the $l$-dependence is the same as before.

\begin{figure}[t]
\vskip -1.8in
\begin{center}
\ \includegraphics[width=30pc]{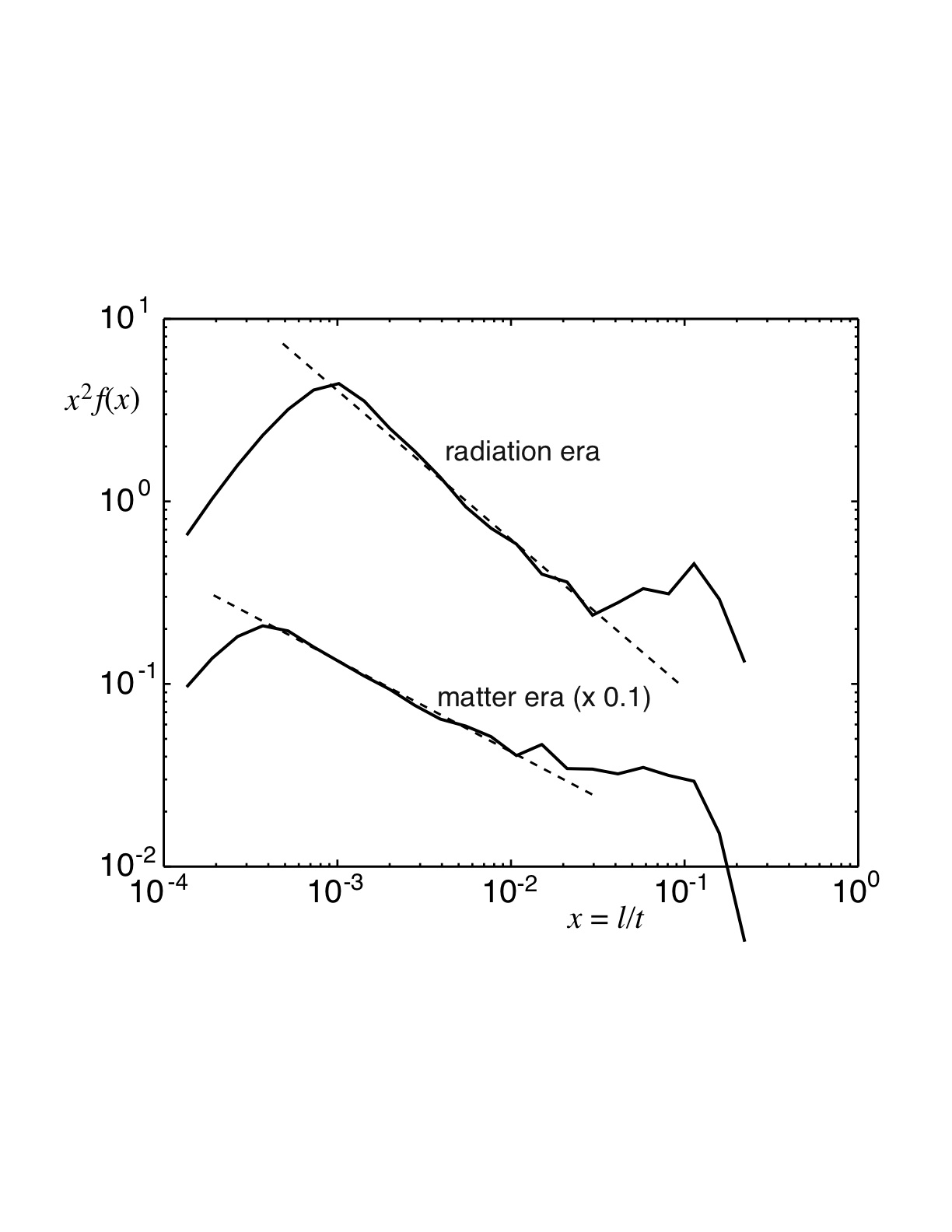}
\end{center}
\vskip -1.5in
\caption{Mean values of $x^2 df(x)$ in the radiation era (upper bold curve) and matter era (lower bold curve, reduced by a factor of 10 for clarity), from Ref.~\cite{Olum:2006ix}.  The dashed lines show the slope predicted by our model.}
\label{dist}
\end{figure}

Fig.~\ref{dist} shows the results of Ref.~\cite{Olum:2006ix} for the average value of 
$x^2 f(x)$, where $x = l/t$ and
\begin{equation}
f(x) = 2 t^3 \gamma^2 \frac{d^3{\cal N}}{dl\,du\,dv}
\end{equation}
(here $\gamma$ is a dimensionless constant: the scaling value of the long string length per unit volume is $\gamma^2/t^2$).  The simulations show a power law distribution above the UV cutoff, with exponent that match rather well with our model (dashed line) in both the matter and radiation eras. 
Ref.~\cite{Ringeval:2005kr} also finds power law distributions for $x^2 f(x)$, with exponents
\begin{eqnarray}
{\rm Ref.}\ \cite{Ringeval:2005kr}:&& x^{-1.0}\ {\rm (radiation)}\ ,\quad x^{-0.5}\ {\rm (matter)}\ ,\nonumber\\
{\rm our\ model}:&& x^{-0.8}\ {\rm (radiation)}\ ,\quad x^{-0.5}\ {\rm (matter)}\ .
\end{eqnarray}
The simulations~\cite{Ringeval:2005kr} and~\cite{Olum:2006ix} thus agree quite well with each other, both matching our model well in the matter era and giving a slightly steeper distribution in the radiation era.

We have found that fragmentation is not necessary to produce the small loops, but it still raises a puzzle.  If we were to apply our same calculation to study fragmentation of these small loops, we would find again a strong tendency to fragment into loops at the UV cutoff scale, giving a sharp peak there rather than a power law distribution.  We believe that the point is that the short distance structure on these small loops is not that typical of the long strings, but is dominated by two or more large kinks moving in each direction.  One kink in each direction forms with the loop, and the remainder are present in the pre-existing distribution.  We have noted in Ref.~\cite{Polchinski:2006ee} that the tails of the small scale structure distribution are not gaussian but are dominated by a single large kink (again, this does not alter the scaling with $l$).  Thus these kinks are likely associated with the tail of the loop production function.
If so, the functions  ${\bf p}_\pm$ on the small loops are dominated by large jumps, from the kinks, and will be unlikely to have cusps where smaller loops can form. 

It may be interesting to consider a renormalization group (RG) analysis, where the separation between the long-distance and short-distance configurations is taken at a running length scale $\tilde L$.  Here we just note a few scaling properties.  Taking the derivative of the two-point function, one finds that at scale $\tilde L$ the rate of change of the tangent vectors is of order $(\tilde L/t)^{-1+ \chi}$.  Thus the typical cusp velocities~${\bf V}_\pm$ are increased by this factor, and the size of each cusp correspondingly decreased.  At the same time, the lengths of the classical curves ${\bf l}_\pm$ are increased by this factor, and so the number of intersections between them increases as $(\tilde L/t)^{-2+ 2\chi}$.  This gives another way to understand the loop production functions: it is plausible that the loop production per logarithmic scale $l d{\cal N}/dl$ scales as the number of cusps with $\tilde L \sim l$.  This RG approach may be useful for understanding the spatial distribution of loop production, and correlations between loops. 

In conclusion, it appears that numerical and analytic methods are converging on a firm picture of the loop production, with one peak near the horizon scale and one near the UV cutoff, with perhaps the larger fraction of string in small loops.\footnote{We are following Refs.~\cite{Siemens:2001dx,Polchinski:2007rg} in putting the UV cutoff at the gravitational radiation scale.  M. Hindmarsh (private communication) has argued that gravitational radiation may be less efficient than believed, in which case the cutoff may be at a smaller scale.}
In Ref.~\cite{Polchinski:2007qc} some observational consequences of this have been discussed.

{\it Note added:}  After the first version of this paper was posted, Ref.~\cite{Vanchurin:2007ee} appeared with a related title but different conclusions.  Ref.~\cite{Vanchurin:2007ee} contains several calculational errors that invalidate its conclusions. In particular, its quantity $\lambda_3(\Delta)$ is actually divergent, as follows immediately from the definition and the fact that $b < 2$ (note that the sign of $c$ must also be corrected).  This divergence has a simple origin.  At any kink, ${\bf a''}$ will have a delta-function contribution, so the RMS value will necessarily diverge.  Omitting this divergence essentially ignores the kinks, which of course gives an incorrect loop distribution.

\section*{Acknowledgments}

We thank Mark Hindmarsh, Carlos Martins, Ken Olum, Paul Shellard, Vitaly Vanchurin and Alex Vilenkin for discussions and communications. 
This work was supported in part by NSF grants PHY05-51164 and PHY04-56556.  The work of F.D. is supported by the Swiss National Funds.  J.V.R. acknowledges financial support from {\it Funda\c{c}\~ao para a Ci\^encia e a Tecnologia}, Portugal, through grant SFRH/BD/12241/2003.%

\end{document}